\documentclass[conference]{IEEEtran}
\usepackage{cite}
\usepackage{graphicx}
\usepackage{amsmath}
\usepackage{listings}
\usepackage{url}
\usepackage[rightcaption]{sidecap}
\usepackage{epstopdf}
\usepackage[caption=false]{subfig}
\usepackage{multirow}
\usepackage{wrapfig}
\usepackage{array}
\usepackage{algorithm,algorithmic}
\usepackage{tikz}
\usepackage[shortlabels]{enumitem}
\usepackage{amsthm}
\usepackage{svg}

\theoremstyle{definition}

\usepackage{listings}
\usepackage{xcolor}

\definecolor{codegreen}{rgb}{0,0.6,0}
\definecolor{codegray}{rgb}{0.5,0.5,0.5}
\definecolor{codepurple}{rgb}{0.58,0,0.82}
\definecolor{backcolour}{rgb}{0.95,0.95,0.92}

\lstdefinestyle{mystyle}{
  backgroundcolor=\color{backcolour},   commentstyle=\color{codegreen},
  keywordstyle=\color{magenta},
  numberstyle=\tiny\color{codegray},
  stringstyle=\color{codepurple},
  basicstyle=\ttfamily\footnotesize,
  breakatwhitespace=false,         
  breaklines=true,                 
  captionpos=b,                    
  keepspaces=true,                 
  numbers=left,                    
  numbersep=5pt,                  
  showspaces=false,                
  showstringspaces=false,
  showtabs=false,                  
  tabsize=2
}

\ifCLASSINFOpdf
\else
\fi

\begin{document}
%


\title{Architecture Support for FPGA Multi-tenancy in the Cloud}


%
\author{
\IEEEauthorblockN{
Joel Mandebi Mbongue,
Alex Shuping,
Pankaj Bhowmik, 
Christophe Bobda}

\IEEEauthorblockA{ ECE Department, University of Florida, Gainesville FL, USA\\
Email: (jmandebimbongue, alexandershuping, pankajbhowmik)@ufl.edu, cbobda@ece.ufl.edu}
}


\maketitle
\vspace{-35pt}
\begin{abstract}
Cloud deployments now increasingly provision FPGA accelerators as part of virtual instances. While FPGAs are still essentially single-tenant, the growing demand for hardware acceleration will inevitably lead to the need for methods and architectures supporting FPGA multi-tenancy. In this paper, we propose an architecture supporting space-sharing of FPGA devices among multiple tenants in the cloud. The proposed architecture implements a network-on-chip (NoC) designed for fast data movement and low hardware footprint. Prototyping the proposed architecture on a Xilinx Virtex Ultrascale+ demonstrated near specification maximum frequency for on-chip data movement and high throughput in virtual instance access to hardware accelerators. We demonstrate similar performance compared to single-tenant deployment while increasing FPGA utilization ( we achieved 6$\times$ higher FPGA utilization with our case study), which is one of the major goals of virtualization. Overall, our NoC interconnect achieved about 2$\times$ higher maximum frequency than the state-of-the-art and a bandwidth of 25.6 Gbps.
\end{abstract}

\begin{IEEEkeywords}
Cloud, Field-Programmable Gate Array, Network-on-chip, Multi-tenancy, Elasticity
\end{IEEEkeywords}

\section{Introduction}
 In recent years, Field-Programmable Gate Arrays (FPGAs) have increasingly been deployed in public cloud infrastructures provided by several technology companies such as Amazon and Alibaba \cite{amazon, alibaba}. Developers can now take advantage of a rich library of pre-built hardware accelerators or implement custom features without purchasing FPGA boards, managing expensive licenses, setting up the operational infrastructure, and being obliged to work from a specific location. Though FPGAs in the cloud opens to a broader access to reconfigurable hardware, current commercial  cloud systems have highlighted the lack of primitives and support allowing multiple workloads to space-share a single device. This could result in expensive utilization cost. An instance without FPGA can be about 8.5$\times$ cheaper than an equivalent with FPGA \cite{f1,ec2_pricing}. Another issue is the waste of resource.  In fact, FPGA devices most often gather more elements than what user workloads would typically need when considering the millions of components present in high-end FPGAs. As example, the Xilinx Virtex UltraScale+ FPGA deployed within Amazon F1 instances contains approximately 2.5 millions logic elements, 6800 DSP slices, and 75MB of BRAM  \cite{ultraScale}.

Because the capacity of integration in FPGA technology continuously increases as some devices now achieve 9 millions of logic cells  \cite{vu19p}, 
we believe that single-tenant FPGA use in the cloud may soon be unsuited. It then becomes necessary to explore approaches to enable FPGA multi-tenancy.  The National Institute of Standards and Technology (NIST) proposed several characteristics of cloud infrastructure among which is \textit{resource pooling} and \textit{rapid elasticity} \cite{mell2011nist}. The \textit{resource pooling} refers to the consolidation of resources (storage, processing, memory, etc) to serve users in a multi-tenant model. On the other hand, the \textit{rapid elasticity} consists in allowing the provision and release of resources. It also encompasses scaling services with the demand. Extending these concepts to cloud FPGAs could summarize in being able to run multiple accelerators on a single device simultaneously, and enable the allocation of additional FPGA resources at run-time.

In this paper, we propose an approach for FPGA virtualization in cloud infrastructure that addresses \textit{resource pooling} and \textit{rapid elasticity}. Since the elasticity assumes that resources can be acquired and ultimately released, we focus our study on FPGA sharing in the space domain. In order to allow logically isolated workloads to share a single device, we start by dividing FPGAs into disjoint regions. The regions are then interfaced to a network-on-chip (NoC) interconnect that allows extending the hardware domain of a task. Basically, a hardware task that is deployed over multiple FPGA regions can be seen as an application with several sub-functions that can communicate through the NoC, each one deployed in a separate location. Our contribution therefore includes:
\begin{enumerate}
   \item Concept of elastic and multi-tenant FPGAs in the cloud. 
    \item A soft NoC for efficient on-chip communication between hardware accelerators. We optimize the NoC architecture considering the cloud needs, and provide a solution that can move data at about 1GHz for data width between 64 and 256 bits.
    \item A case study on FPGA multi-tenancy and elasticity. It shows through a practical example the necessity of space-sharing, and illustrates the advantage of on-chip communication support for efficient elasticity.
    
\end{enumerate}

In the rest of the paper, section  \ref{sec:rel_work} reviews recent research. Then, section \ref{sec:fpga_multitenant_elastic} provides some background definitions. Next, section \ref{sec:proposed_noc} describes the components of the proposed soft NoC. Finally, section \ref{sec:experiments} shows some experimental results and section \ref{sec:conclusion} concludes the paper.

\section{Related Work}
\label{sec:rel_work}
\subsection{FPGAs in the cloud}
\label{subsec:fpga_cloud}
FPGA virtualization in the cloud has been discussed recently in several studies.
Some contributions present solutions to the temporal allocation of FPGA kernels \cite{ tarafdar2019building, dai2014online, zhang2019computer, al2019cloud}. They essentially explore techniques to successively allocate full or partial FPGAs to tenants over time. 
Other research illustrated architectures implementing spatial FPGA sharing by exposing FPGA regions labeled "\textit{virtual FPGAs}" to cloud tenants. For instance, some architectures divide each physical FPGA into several locations that can be allocated to virtual instances (VI) \cite{chen2014enabling,byma2014fpgas,weerasinghe2015enabling}. Partial reconfiguration is then used for runtime update of VI's hardware kernels. Yet, the FPGA access remains mostly limited either by not allowing user custom designs but pre-built hardware functions, and/or not supporting direct on-chip communication. This restriction imposes middleware copy for data movement between accelerators.  To minimize the data movement overhead, an on-chip interconnect can be used between virtualized hardware regions \cite{mbongue2018fpga,mbongue2018fpgavirt}. Vaishnav et al. implement elasticity on cloud workloads by scheduling user jobs as they arrive \cite{vaishnav2018resource}. Based on a list of waiting jobs and their needs, they use partial reconfiguration to repurpose the FPGA. Other work discuss security challenges of FPGA deployment in the cloud \cite{hategekimana2018secure}. This aspect is out of the scope of this work.

\subsection{Network-on-Chip}
\label{subsec:noc}
Network-on-chips have emerged as a solution to the lack of scalability of point-to-point links and buses. Several soft-NoCs implemented on FPGA have been proposed in the literature. In the FLexiTASK NoC, high-radix routers reduce the network diameter \cite{mandebi2018flexitask}.
Schelle et al. explore NoC performance when modifying parameters such as the size of the network and the presence or not of virtual channels (VC) \cite{schelle2008exploring}. They showed for instance that VCs can lead to about $5\times$ increase in resource utilization, but allow higher throughput. They concluded that more logic should be spent on the NoC if the applications are more communication-oriented than compute. CONNECT is a flexible NoC generator for FPGA-based systems that allows the creation of arbitrary topologies \cite{papamichael2012connect}. Its flexibility however results in low $F_{max}$ and high area overhead. 
Hoplite proposes a lightweight and bufferless router architecture that is capable of achieving high bandwidth for single-flit-oriented FPGA designs with low area overhead \cite{Kapre2015}. Our proposed NoC is inspired from Hoplite as we seek to minimize the hardware footprint of the NoC to make more resources available to user designs in the cloud. Maidee et al. present a topology that leverages under-utilized FPGA long wires at the edge of the device for fast data movement \cite{maidee2017linkblaze}. Our proposed topology similarly leverages long wires on FPGA devices. Discussions on hard-NoCs are out of the scope of this work.

\section{FPGA Multi-tenancy and Elasticity}
\label{sec:fpga_multitenant_elastic}
\subsection{Background Concepts}
\label{subsec:background_concept}

The virtualization of computing components such as CPUs is well investigated and basically consists in running virtual CPU instructions on a physical processor. While instructions on a CPU can execute independently, FPGAs implement circuits that depend on the physical architecture of the underlying hardware. This therefore requires reserving some FPGA regions to place run-time workloads as opposed to scheduling instructions as in the case of CPUs. In the context of this work, we define \textbf{FPGA Multi-tenancy} as  the capability of
space-sharing the physical area of a device between hardware accelerators from different cloud users. The placement of several hardware kernels consequently imposes splitting the FPGA into non-overlapping areas that we call "\textit{virtual region}" (VR). VRs represent the unit of virtualized FPGA resource in the cloud. We consider the \textbf{FPGA Elasticity} as a feature that enables assigning additional unit of FPGA virtualization to already deployed tasks run-time with support for on-chip sub-function communication. 

\subsection{Cloud Virtualization Model}
\label{subsec:virt_model}
\begin{figure}[h]
\vspace{-6pt}
\flushleft
\includegraphics[width=0.495\textwidth]{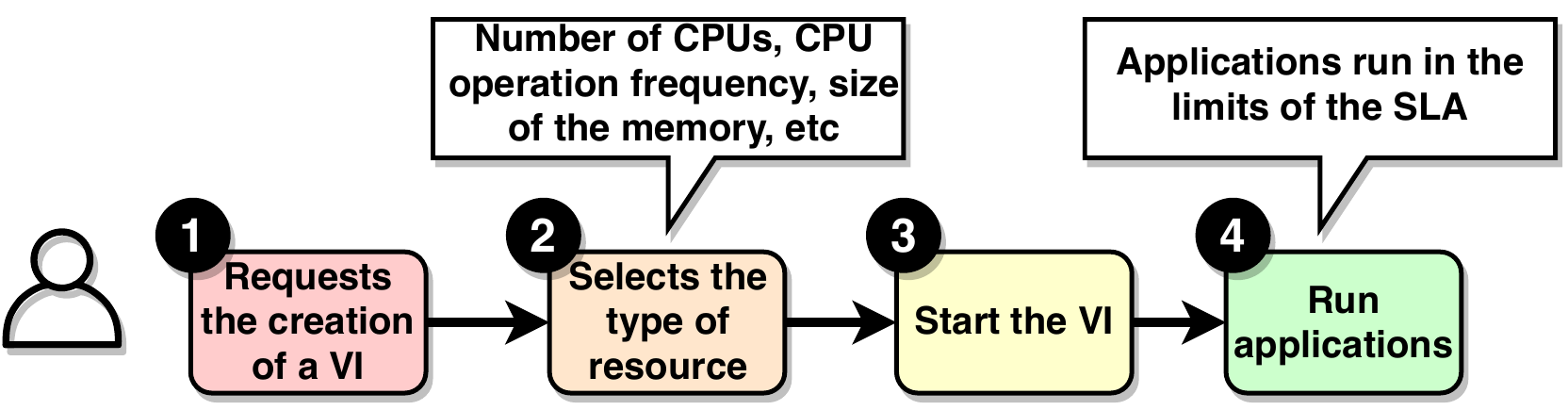}
\caption{Regular Virtual Instance Creation Flow}
\label{fig:vi_creation} 
 \vspace{-6pt}
\end{figure}

In this work, we consider the virtual resource access flow illustrated in Figure \ref{fig:vi_creation}. It starts with a user request to the cloud provider for setting up a virtual instance (VI). The user selects the resources to attach to the desired VI and can start running applications.  
Tasks can run as long as they do not violate the Service-Level Agreement (SLA). For instance, if a VI is set up with a disk of 1TB, it will not be possible to store more data until requesting additional storage. This flow is generally adopted in cloud infrastructures delivering VI. Our work seeks to enable selecting FPGA unit of virtualization as part of VIs. 
The size and shape of each VR is left to the cloud provider's choice just as they decide what unit of memory, storage, and processing they offer in their VI flavors. Since the amount of logic in an FPGA is finite, the same goes for the area of each VR. In consequence, the designs that are larger than a VR will be divided into modules by the user just as it would be the case if a design was bigger than an entire device. Next, the user will place a request for additional FPGA unit of virtualization. Because the two user regions will eventually exchange data, we propose to provide an efficient NoC interconnect as part of the Shell on FPGA. The NoC will also enable extending deployed workloads with additional functions at different VR. Our concept of elasticity differs from that of Vaishnav et al. \cite{vaishnav2018resource}, as we consider a model in which users fully control (run-time programming through partial reconfiguration) units of FPGA assigned to their domains by the cloud infrastructure. 
\begin{figure*}[h]
\captionsetup[subfloat]{farskip=2pt,captionskip=1pt}
	\centering
   
    \subfloat[Buffered Bidirectional Router ]{
		\label{fig:bufferedRouter}
		\includegraphics[width=.2\textwidth]{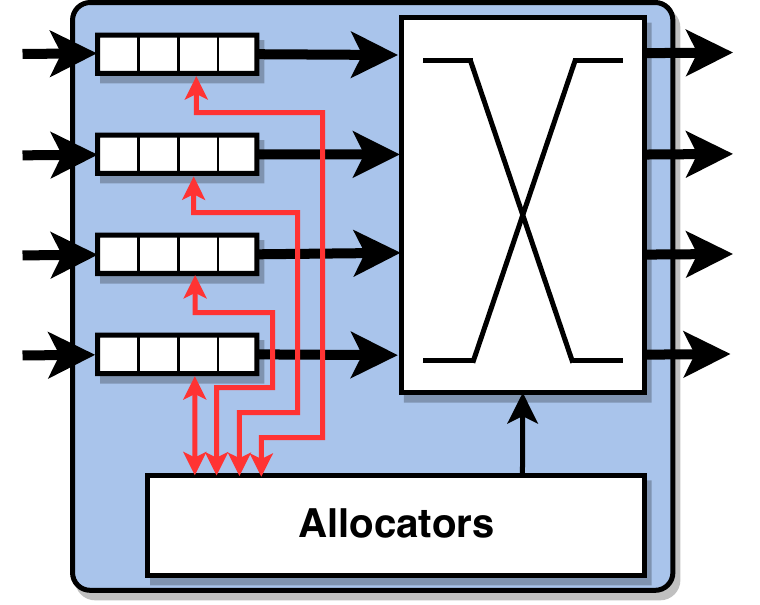}%
	}		
	\hspace{0.4cm}
	\subfloat[Bufferless Bidirectional Router and VR Architecture]{
		\label{fig:bufferLessRouter}
		\includegraphics[width=.65\textwidth]{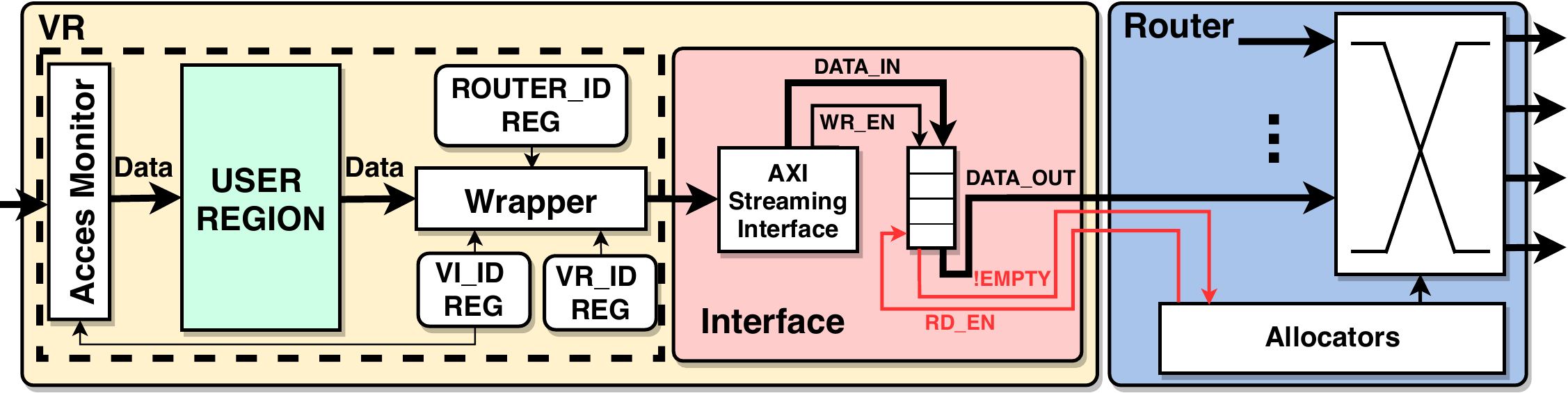}	}
		
		\caption{Router optimization for $F_{max}$ improvement and area reduction.}
 	\label{fig:displyaOnSmartphone} 
  	\vspace{-6pt}
\end{figure*}

\section{Proposed Network-On-Chip Architecture}
\label{sec:proposed_noc}
In order to efficiently implement the \textit{resource pooling}, we seek to maximize the number of concurrent workloads that can be deployed on a single device. In other words, we attempt to minimize the amount of resources consumed by the shell (NoC and IO controllers). We will however focus on optimizing the NoC architecture. Further, for fast data movement between VRs, the NoC should achieve device specification $F_{max}$, which is related to decreasing the number of LUTs on the datapaths. 

 \subsection{Proposed Topology}
 \label{subsec:topology}
\begin{figure}[h]
\vspace{-8pt}
\captionsetup[subfloat]{farskip=2pt,captionskip=1pt}
	\centering
   
    \subfloat[]{
		\label{fig:traditionalMesh}
		\includegraphics[width=.16\textwidth]{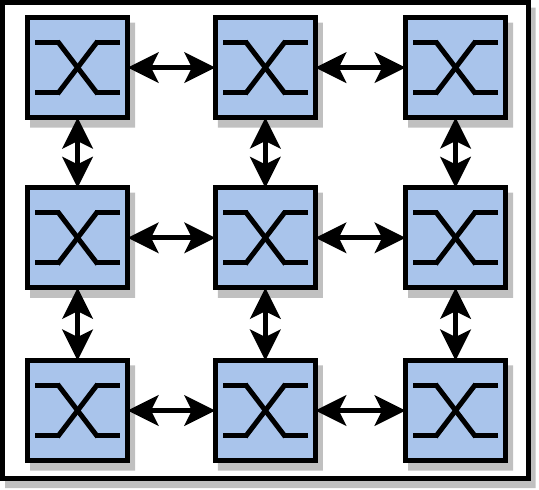}%
	}		
	\hspace{0.0cm}
	\subfloat[]{
		\label{fig:ourMeshTopology}
		\includegraphics[width=.195\textwidth]{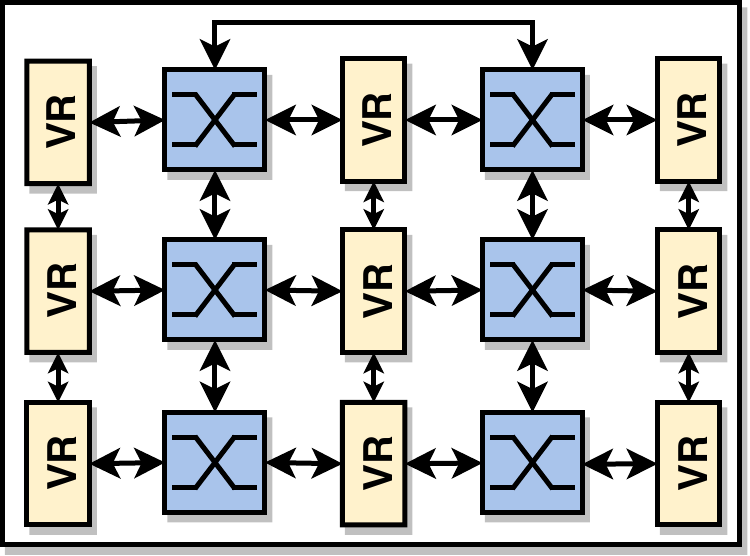}	}

		\caption{(a) Traditional 2D Bidirectional Mesh Architecture. (b) Our proposed NoC topology. It reduces the radix of routers and enables direct inter-VR communication.}
 	\label{fig:noc_topology} 
     \vspace{-6pt}
\end{figure}

 While there are several topologies such as ring, star, hypercube, etc; we consider a 2D Mesh style for our NoC architecture. Mesh topologies usually feature processing elements (PE) with a network interface attached to a router. Architectures implementing a 2D mesh typically have routers with 5 interfaces (4 interfaces to communicate with adjacent routers and 1 interface attached to a PE). Figure \ref{fig:traditionalMesh} illustrates a general view of a 3$\times$3 2D mesh. Mesh topologies have two defects in term of the FPGA logic needed for each router and the overall communication latency. (1) A smaller network diameter is tightly coupled to a larger router radix (number of IO ports of the router). This allows reaching destinations in a few hops from any source and possibly reduce communication overhead. However, crossbars and allocators are well known to grow quadratically in logic with the radix of the routers, resulting in substantial routing delays, lower operating frequency, and higher area and power consumption. (2) In a mesh, each router serves a single PE. This means that any communication between PEs requires a minimum of 2 hops, each router introducing potential delays depending on the traffic. Because we target lower resource utilization, high frequency of operation and low communication latency, we propose the topology illustrated in Figure \ref{fig:ourMeshTopology}. It is a 3$\times$3 mesh in which routers are connected to VRs. It implements a topology in which routers have at most 4 ports. As opposed to a regular mesh, each router is connected to 2 VRs, which decreases the hops. In order to keep the radix of routers to 4 with 2 VRs connected, we reduce the dimension of the routing. Packets are either pushed up/down or injected into the VRs. We also enable direct communication links between VRs, which allows offloading routers and streaming data every clock cycle between adjacent workloads.

 Depending on the width of a device and the size of the VRs, the topology can be deployed in three different flavors: \textbf{(1) Single-Column:} in which the routers are lined up vertically on a few columns of configurable logic blocks (CLB). \textbf{(2) Double-Column:} it uses two columns of routers as in Figure \ref{fig:ourMeshTopology}. In this mode, underutilized wires at the edge of the device are used to connect the two columns of routers. In fact, unless specified placement constraints, vendor tools tend to place and route designs closer to the center of the chip. By using wires at the edge, we take advantage of commonly wasted FPGA resources to provision additional VRs. \textbf{(3) Multi-Column:} it extends the previous mode with additional columns of routers and is suitable for wider devices.

 We leverage architecture optimization in high-end FPGAs to maximize the NoC operating frequency while reducing the area and power consumption. For instance, UltraScale devices are arranged in a column-and-grid layout of clock regions that are 60 CLBs height. A CLB contains eight 6-LUTs and 16 flip-flops. This high capacity of integration allows packing the NoC routers over a few CLBs ($<$\%1 of the chip). In addition, rapid signal transmission is made possible by the abundance of switches and long wires spanning 16 CLBs \cite{ultrascale_overview}. With fabric switches connecting large datapaths, the NoC can implement high frequency wide buses. We use placement constraints to force NoC into specific areas of the chip and prevent CAD tools from using more CLBs than necessary. Next, we constrain routing within the boundaries of the NoC allocated areas, freeing up more resources for user designs. Our NoC implementation uses the AXI4 interfaces for standardization.
 
 Though our topology may lead to higher hops compared to a traditional mesh in some cases, its higher connectivity between VRs offers more flexible placement options.

 
  \subsection{Router Component}
  \label{sec:router}
  \subsubsection{Architecture}
 In this section, we discuss design choices and optimization on the router's internal architecture.
 
 We start with the typical bidirectional router architecture that is presented in Figure \ref{fig:bufferedRouter}. The \textit{Input Buffers} serve two purposes: (1) enabling minimized event of metastability between VR and router clock domains. (2) Temporary data storage when the destination is not ready.  In order to forward traffic to the destination, each router implements a \textit{Crossbar Matrix} that connects input and output channels, and allows parallel data streaming. We optimize the size of the crossbar by removing unnecessary muxes. In fact, if we consider that we have $n$ inputs and $m$ outputs, each of the output lines only needs $n-1$ switches since it is not the case that a VR will send data to itself. Each router therefore has $(n-1) \times m$ switches in the crossbar instead of $n \times m$. With 4-port routers,  each line in the crossbar thus multiplexes three entries. 
 In our topology, the first and last routers only need three interfaces (see Figure \ref{fig:ourMeshTopology}). This is simply a consequence of the absence of a fourth component to attach. Because one of the goals is to keep a low hardware footprint, we implement a 3-port version of the router. This reduces the number of switches to 2 on each line of the crossbar. It also gives cloud providers the flexibility to assemble the topology that meets their needs by combining routers with 3 and 4 interfaces.
 
\begin{figure}[]
\centering
\vspace{-10pt}
\hspace{-0.8cm}
\begin{minipage}[t]{5cm}
  \centering
  \includegraphics[scale=0.42]{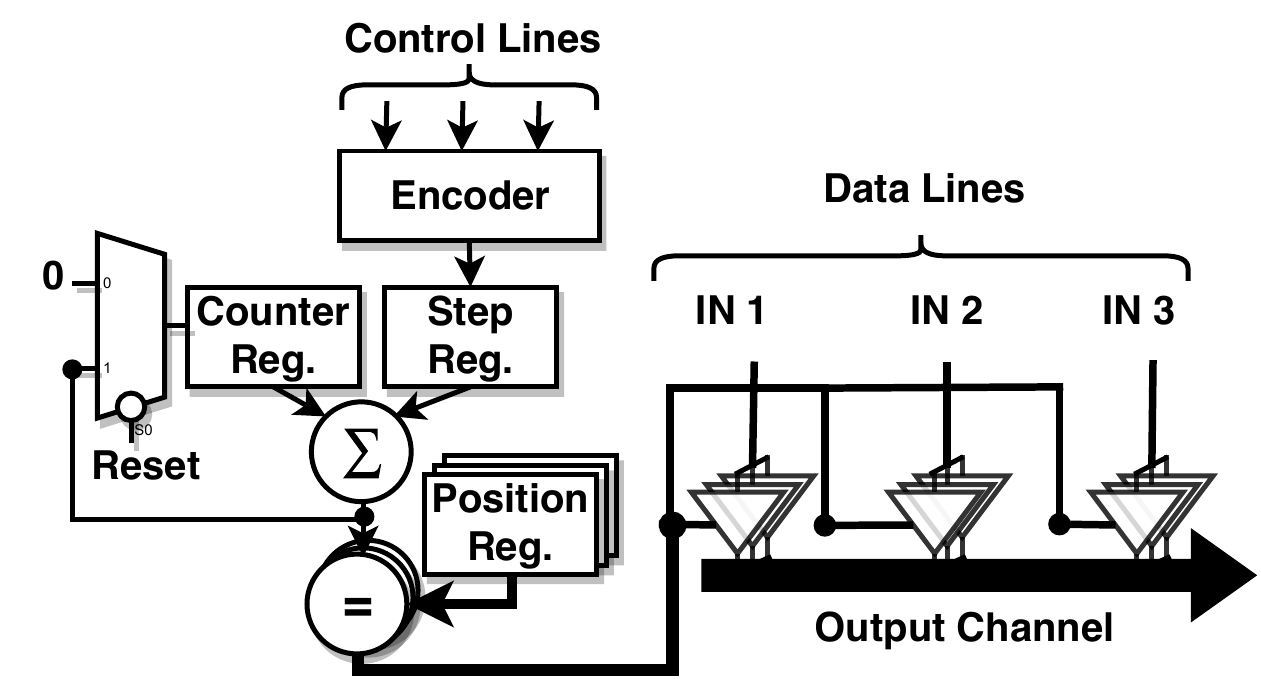}
  \caption{Mutual Exclusion Logic}
  \label{fig:mutex_logic}
\end{minipage}
\hspace{0.4cm}
\begin{minipage}[t]{3cm}
  \centering
  \includegraphics[scale=0.6]{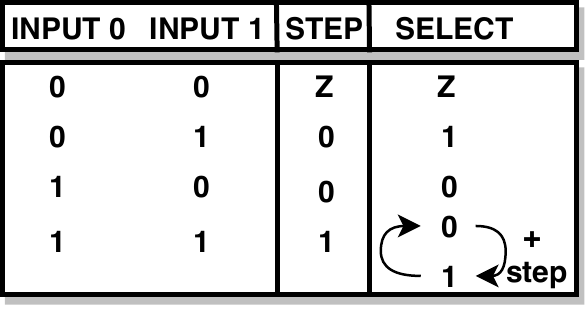}
  \caption{2-Input Encoder}
  \label{fig:encoder}
\end{minipage}
\vspace{-5pt}
\end{figure}

Kapre et al. observed that buffers can increase router resources by $20\%-40\%$, which comes at the cost of area, delay, and power \cite{Kapre2015}. As in Hoplite, we therefore implement bufferless routers as illustrated in Figure \ref{fig:bufferLessRouter}. We remove the buffers from the routers and keep data within VRs until the routers is ready process the packets. The \textit{Allocators} are responsible for loading the data into the crossbar. Each allocator monitors a specific channel of the crossbar and implements a 3-way handshake protocol that works as follows: (1) The VR lets the allocator know that data is available through the buffer "EMPTY" signal. (2) When the crossbar is ready, the allocator pulls the data by asserting the "RD\_EN" signal. (3) The data is loaded in the crossbar.  Each allocator is also responsible for mutual exclusion between packets that pass through the same crossbar output channels. The purpose is to make sure that only one packet crosses an output channel at a time. Figure  \ref{fig:mutex_logic} summarizes the mutual exclusion logic. Based on the control lines asserted that signals the presence of incoming packets, an encoder determines the packet that is read in. If there are multiple packets from different sources, one packet is pulled from input interface at a time to establish fairness. Figure \ref{fig:encoder} shows the logic of the encoder.

\begin{figure}[h]
\vspace{-6pt}
\centering
\includegraphics[width=0.3\textwidth]{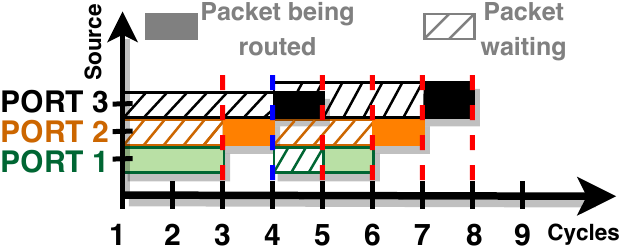}
\caption{Illustration of the mutual exclusion when packets at destination of Port4 of arrive simultaneously from Port1, Port2 and Port3 in a 4-port router.}
\label{fig:collisionManagement} 
\vspace{-6pt}
\end{figure}

 To illustrate the management of mutual exclusion, consider a 4-port router with traffic coming from ports 1, 2 and 3 to port 4. Figure \ref{fig:collisionManagement} summarizes how the \textit{allocator} loads the packets. In cycle 1, there are incoming traffic from the 3 ports. The packets are routed one at a time. In cycle 4, when new data arrives at the 3 input ports, the data is loaded in the same way. From the third cycle, data will simply keep flowing out of the router because the inputs are pipelined.
 
\subsubsection{Routing Procedure} 
In this section, we discuss the routing algorithm implemented in our NoC topology. 

\begin{figure}[h]
\vspace{-6pt}
\centering
\includegraphics[width=0.45\textwidth]{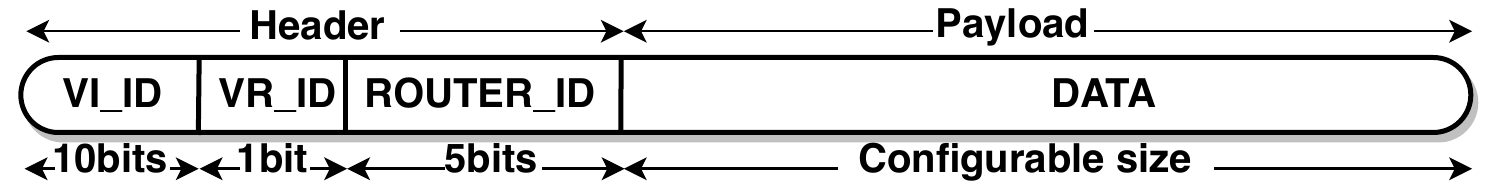}
\caption{Communication Packet Structure}
\label{fig:packet} 
\vspace{-6pt}
\end{figure}

Although we opted for bufferless routers like Hoplite does, we do not implement deflection for two reasons. First, it may lead to unpredictable number of hops. Second, the routers of our topology only route in one dimension. As a result, packets are either injected into one of the VRs connected to the router, or pushed up or down to the next router depending on the destination address. The routing decision is based on the content of each packet header. The packet structure is presented in Figure \ref{fig:packet}. The header has a fixed width of 16 bits and the payload as a configurable size. The header defines the destination of the packets. It is a combination of the VR\_ID and ROUTER\_ID.   
 The VR\_ID is represented on 1 bit. It identifies the VR that is the destination of the packet. Since each router is connected to at most 2 VRs (west and east sides), a VR\_ID that is equal to 0 corresponds to the west VR, and a VR\_ID that is 1 refers to the east VR. The ROUTER\_ID labels the router to which the destination VR is connected. The VI\_ID uniquely identifies the VI to which the packet belongs. It is not actually used in the routing process, but at the VR interface to prevent sending packets to a VR belonging to a different VI. The ROUTER\_ID occupies 5 bits and labels routers with integer values. The VI\_ID occupies 10 bits, which allows handling up to 1024 VIs. Algorithm \ref{algo:routing_algo} summarizes the routing procedure. 

\begin{algorithm}
 \scriptsize
\caption{Packet Routing}
\label{algo:routing_algo}
\begin{algorithmic}[1]
\STATE \textbf{Input:} $incomingPacket,routerId$
\STATE
\FOR{\textbf{each} $incomingPacket$}
   
    \IF{(getRouterID($incomingPacket$) $> routerId$)}
       \STATE forwardToNorth($incomingPacket$);
       \STATE \textbf{goto} Next;
    \ENDIF
   
   \IF{(getRouterID($incomingPacket$) $< routerId$)}
       \STATE forwardToSouth($incomingPacket$);
       \STATE \textbf{goto} Next;
    \ENDIF
    
    \IF{(getVRID($incomingPacket$) $== 0$)}
       \STATE forwardToWest($incomingPacket$);
     \ELSE  
       \STATE forwardToEast($incomingPacket$);
    \ENDIF
   
  \STATE Next: 
 \ENDFOR
\end{algorithmic}
\end{algorithm}
The algorithm first checks the ROUTER\_ID. If the current ROUTER\_ID is greater (resp. smaller) than that of the packet being transmitted, the packet is pushed up (resp. pushed down). If the packet has reached the destination router, the VR\_ID field is checked to determine the VR into which the packet will be injected.

 In the next section, we discuss the structure of the VRs.

   \subsection{Virtual FPGA Region Architecture}
   \label{subsec:vr_architecture}
The architecture of FPGA provisioned regions is illustrated in Figure \ref{fig:bufferLessRouter}. The major component of the VRs is the \textit{USER REGION}. It hosts the cloud user's custom designs and implements the partial reconfiguration paradigm. The VRs also feature an \textit{Access Monitor} which only accepts packets from a specific VI. It removes the packet header and only forwards the payload to the \textit{USER REGION}. The user designs only receive the payloads to prevent malicious application from trying to access resources out of a their domain. Developers are simply provided well-defined interfaces to implement in their design. Next, the cloud infrastructure selects the suitable VR that will host the hardware accelerator. Finally, it programs the design into the \textit{USER REGION} inside the selected VR. At configuration time, the hypervisor edits the content of the VR registers. If the VR communicates with other FPGA regions, the router and VR identifiers of the destination are stored in the \textit{ROUTER\_ID} and \textit{VR\_ID} registers. The VI identifier is also written into the \textit{VI\_ID} register. Whenever a VR is sending a packet out, the \textit{USER REGION} produces the payload that is appended to the header generated in the \textit{Wrapper} module to form a valid packet. Details on algorithms implemented in the hypervisor to efficiently select the VRs to allocate to the VIs are out of the scope of this work. 

\vspace{-4pt}
\section{Experimental Evaluation}
\label{sec:experiments}
\subsection{Evaluation Platform}
\label{subsec:platform}
We prototype the proposed architecture in a cloud configuration comprising two nodes. The first node runs the VIs on OpenStack Stein. It is an all-in-one deployment on a Dell R7415l EMC server running on a 2.09GHz AMD Epyc 7251 CPU with 64GB of memory. The second node hosts the FPGA. It is a Supermicro X10DAx servers with a 3.50GHz Intel Core i7-5930K CPU with 64GB of memory. Both nodes run CentOS-7 with a kernel of version 3.10.0. The servers are connected to a XR700 Nighthawk router operating at a bandwidth of 100Mbps. We use a Xilinx Virtex UltraScale+ FPGA or simply VU9P (xcvu9p-flgb2104-2-i) as testing device. Vivado 2018.2 is used to synthesize, place and route the designs.

\vspace{-3pt}
\subsection{Evaluation Methodolody}
\label{subsec:eval_methodology}
We will proceed in two steps. First, we evaluate the performance benefits of the optimizations discussed in section \ref{sec:proposed_noc}. We assess the performance of our NoC against some metrics such as area, power, maximum frequency, latency, and waiting time. We will also compare our proposed NoC to previous research. Next, we study an example case. We consider a cloud deployment in which multiple VIs are allocated some regions of the FPGA. We do not discuss the VR allocation flow as it is out of the scope of the work, but we analyze the outcome of sharing a device between several tenants by discussing FPGA access time and throughput. We want to demonstrate that sharing the FPGA outcomes in higher FPGA utilization and minimal loss in quality of service (QoS) compared to allocating a whole device to a single tenant. Next, we will compare our results to recent research on FPGA virtualization in the cloud.

\subsection{NoC Evaluation}
\subsubsection{Resource and Power Consumption}
\begin{figure}[]
	\centering
    \subfloat[Our 3-Port Router]{
		\label{fig:3_port}
 		\includegraphics[width=0.18\textwidth]{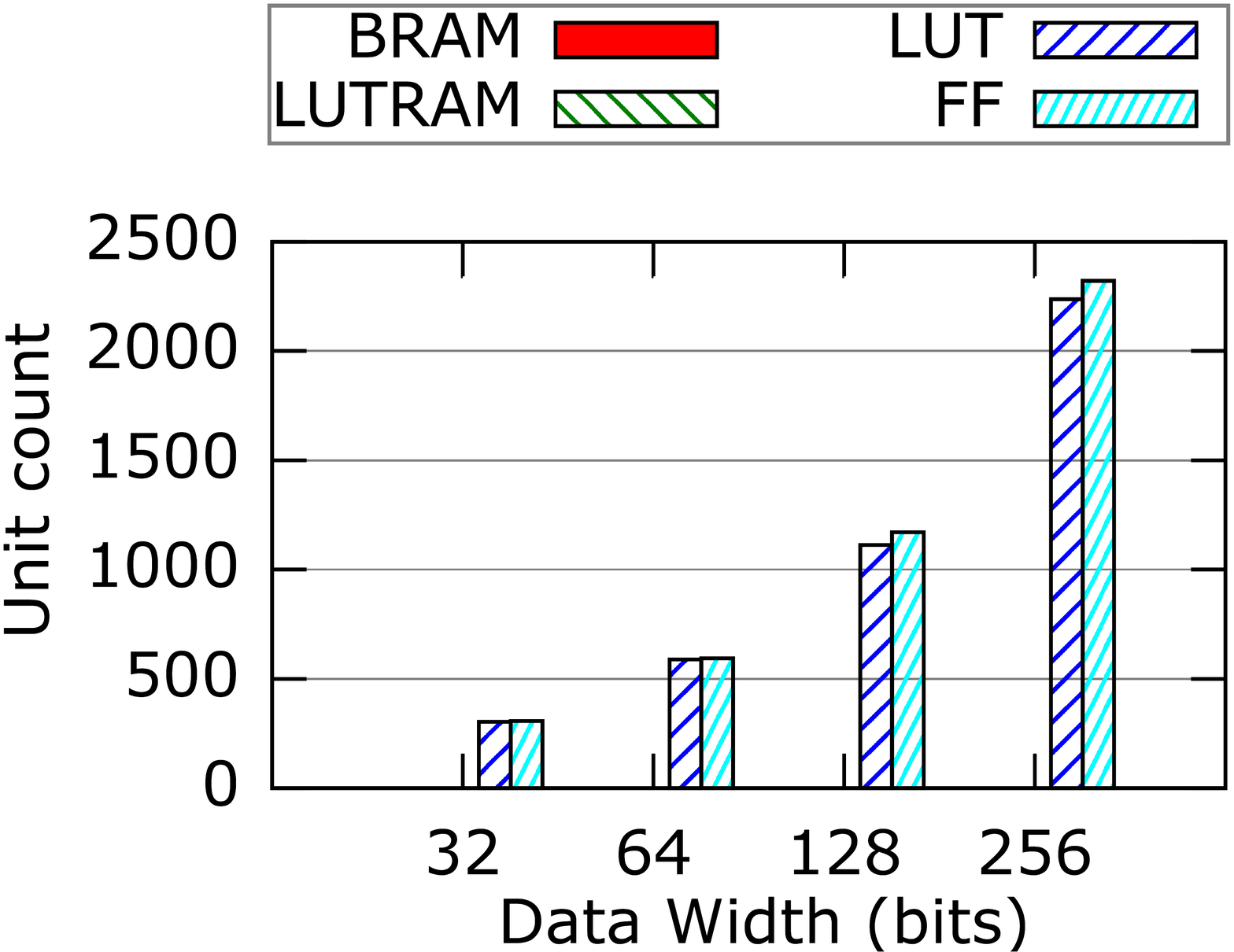}
	}		
	\hspace{0.0cm}
	\subfloat[3-Port Buffered Router]{
		\label{fig:3_port_buffer}
		\includegraphics[width=0.18\textwidth]{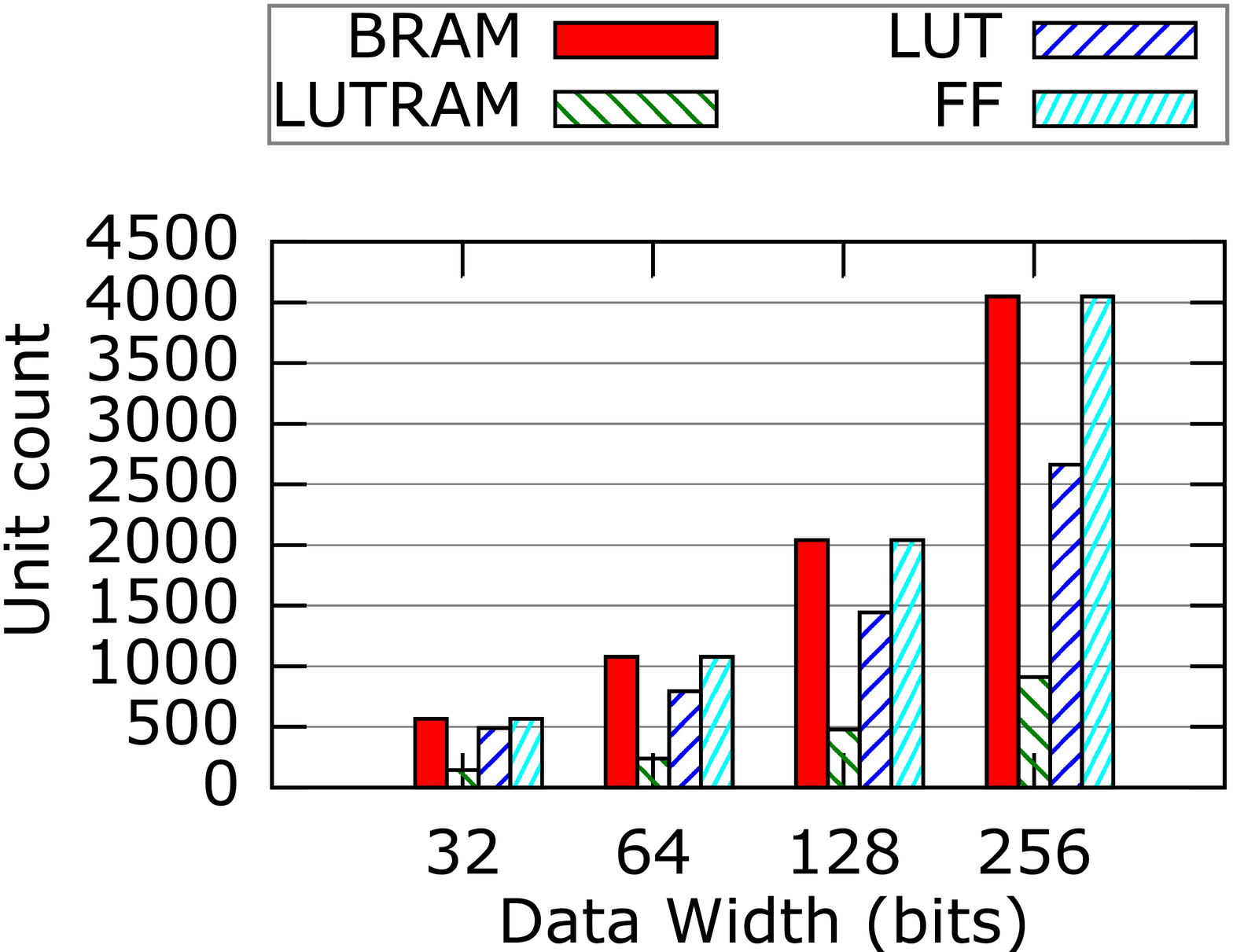}	}
  
    \vspace{-0.2cm}	
		
    \subfloat[ Our 4-Port Router]{
		\label{fig:4_port}
		\includegraphics[width=0.18\textwidth]{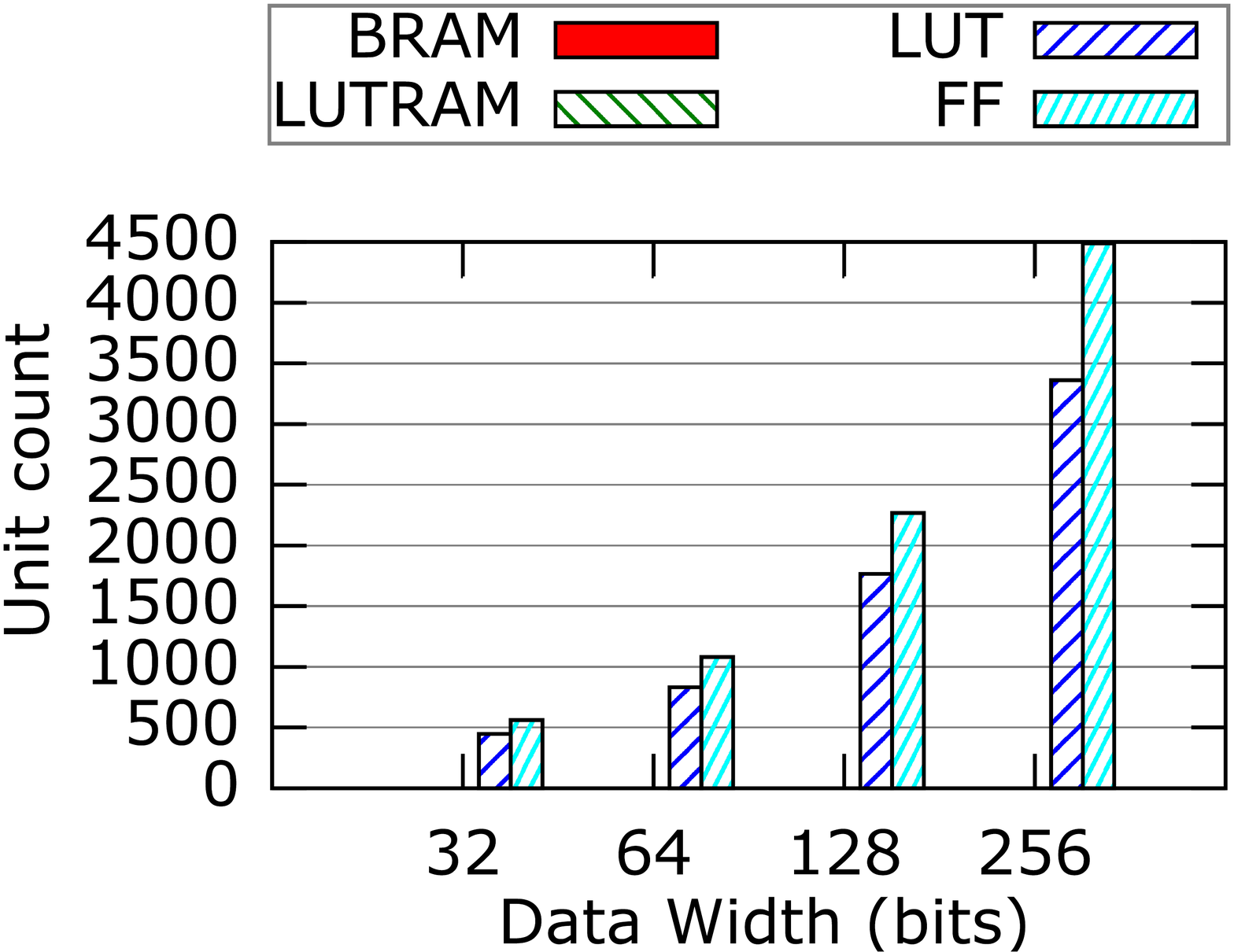}
	}		
	\hspace{0.0cm}
	\subfloat[4-Port Buffered Router]{
		\label{fig:4_port_buffer}
		\includegraphics[width=0.18\textwidth]{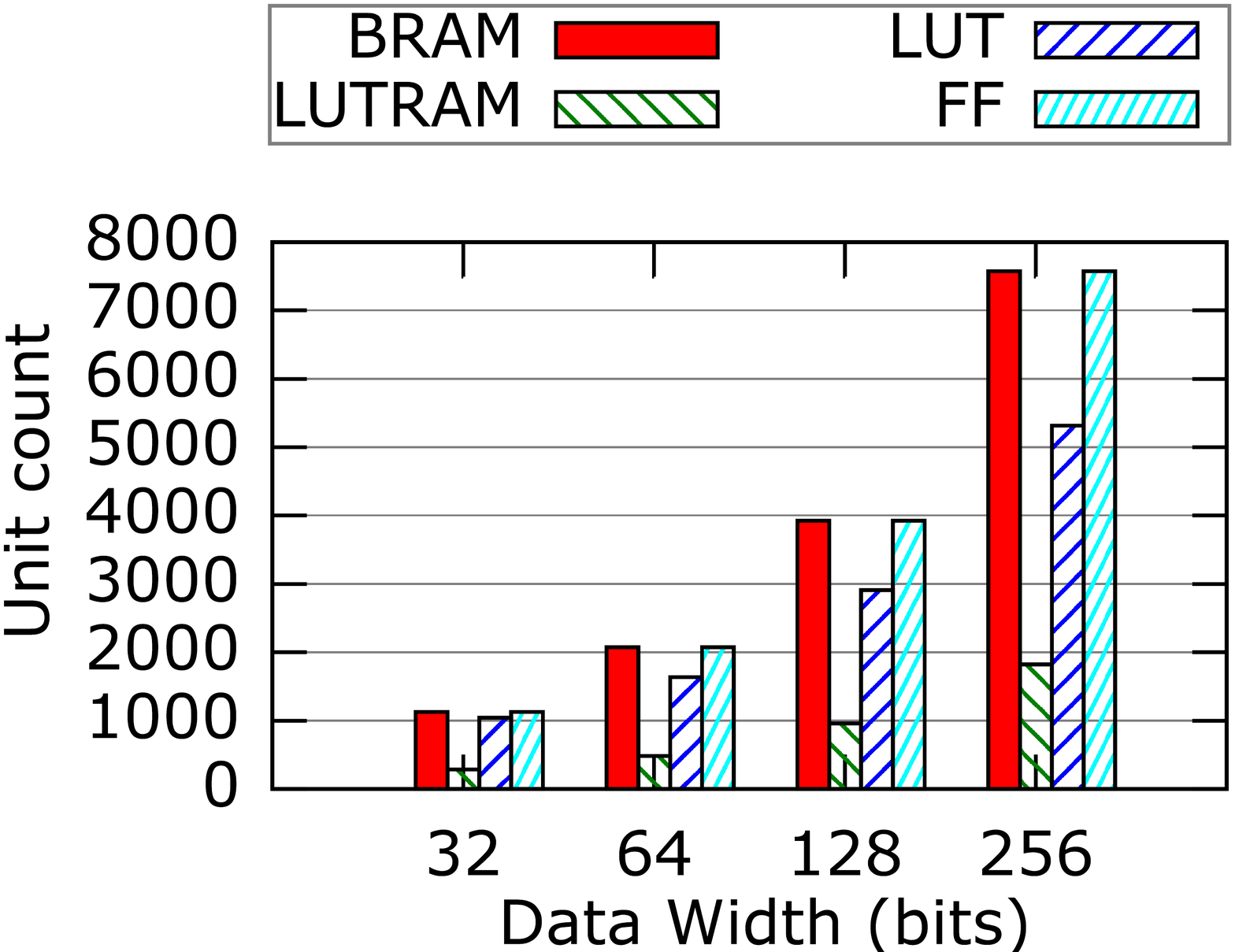}	}
 	\caption{FPGA Resource Utilization of the Router}
 	\label{fig:resourceUtilization} 
 	\vspace{-5pt}
\end{figure}

\begin{figure}[H]
	\centering
	\vspace{-8pt}
    \subfloat[32-bits Routers]{
		\label{fig:luts}
 		\includegraphics[width=0.18\textwidth]{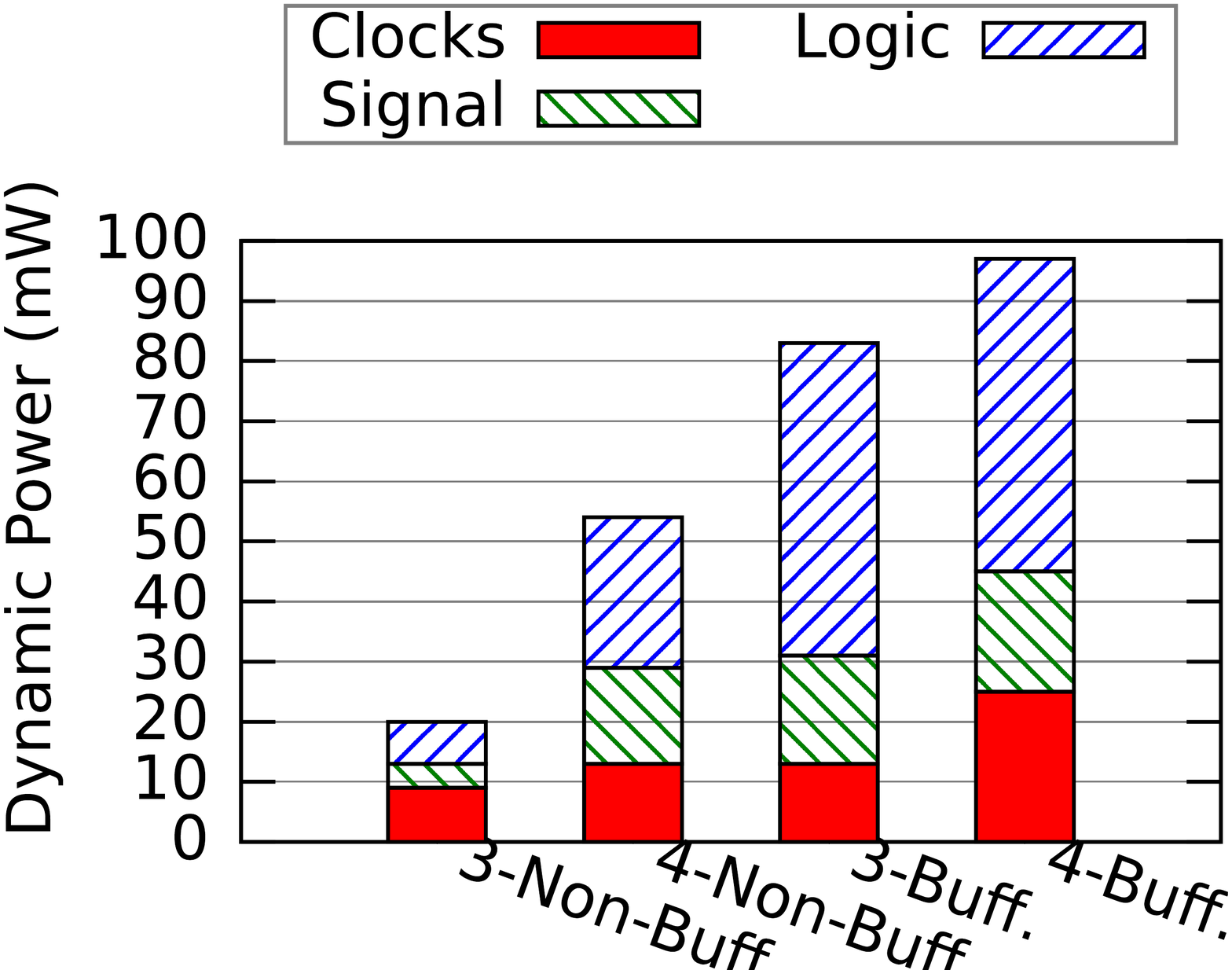}
	}		
	\hspace{0.0cm}
	\subfloat[64-bits Routers]{
		\label{fig:ffs}
		\includegraphics[width=0.18\textwidth]{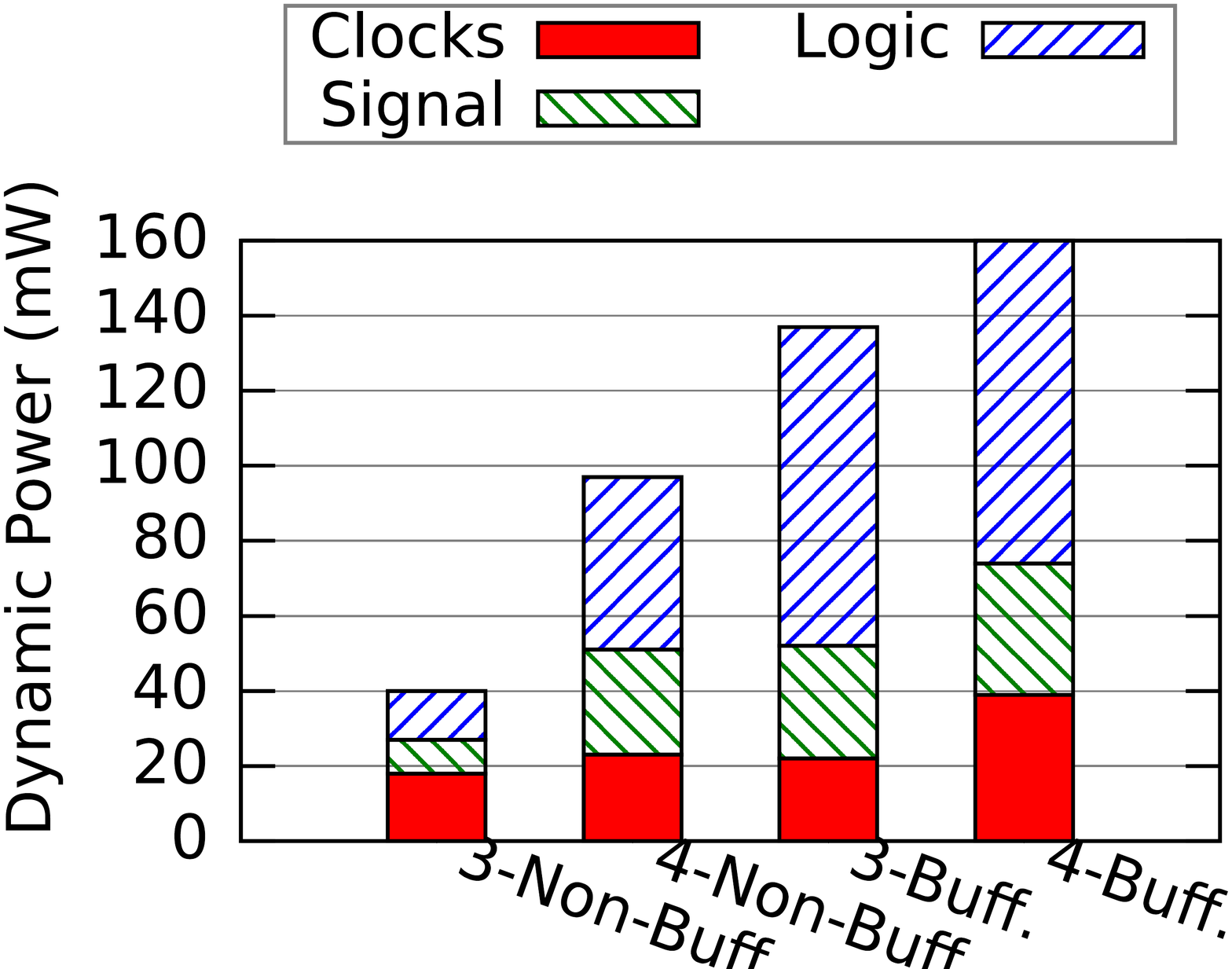}	}
  
    \vspace{-0.2cm}	
		
    \subfloat[128-bits Routers]{
		\label{fig:dsps}
		\includegraphics[width=0.18\textwidth]{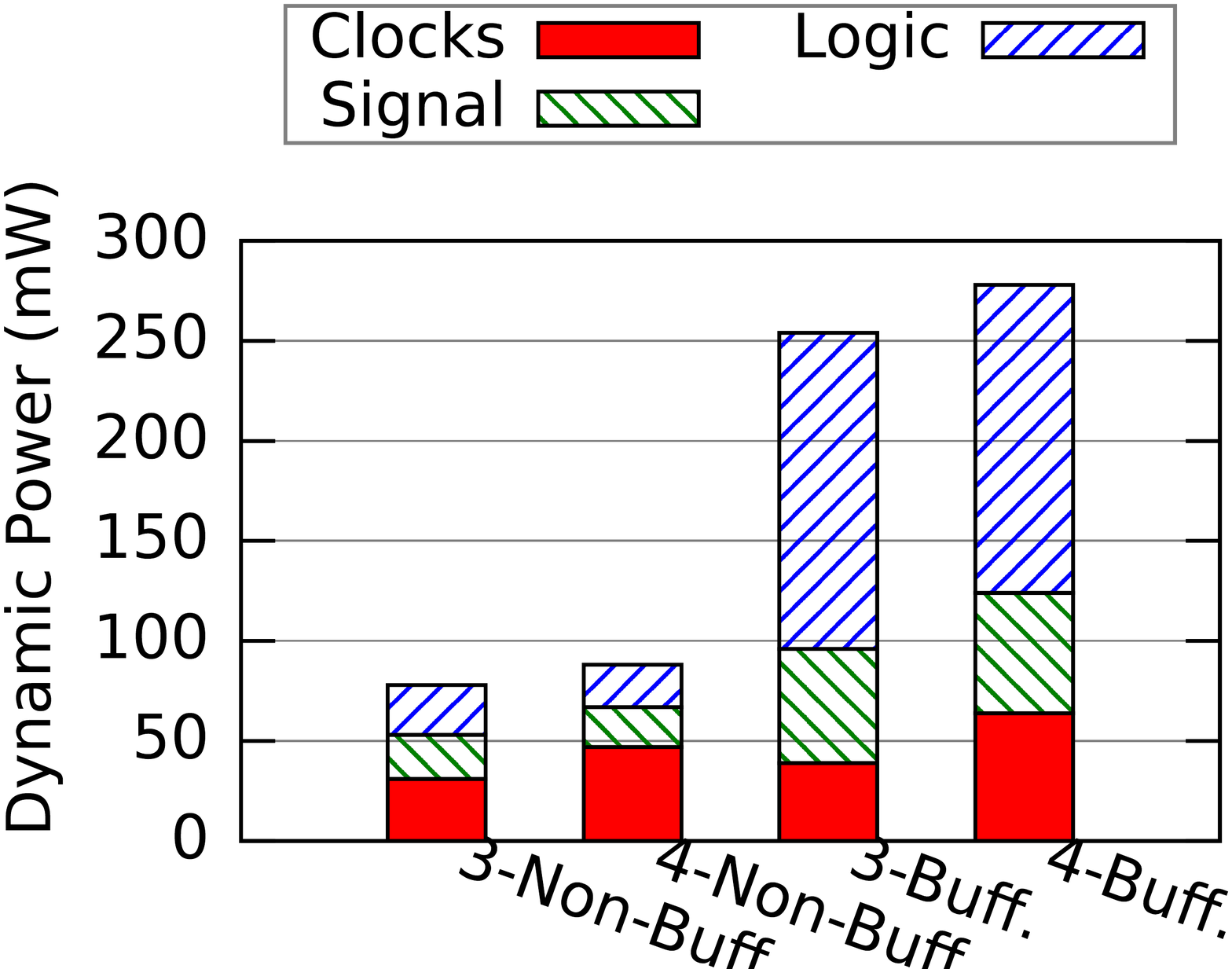}
	}		
	\hspace{0.0cm}
	\subfloat[256-bits Routers]{
		\label{fig:brams}
		\includegraphics[width=0.18\textwidth]{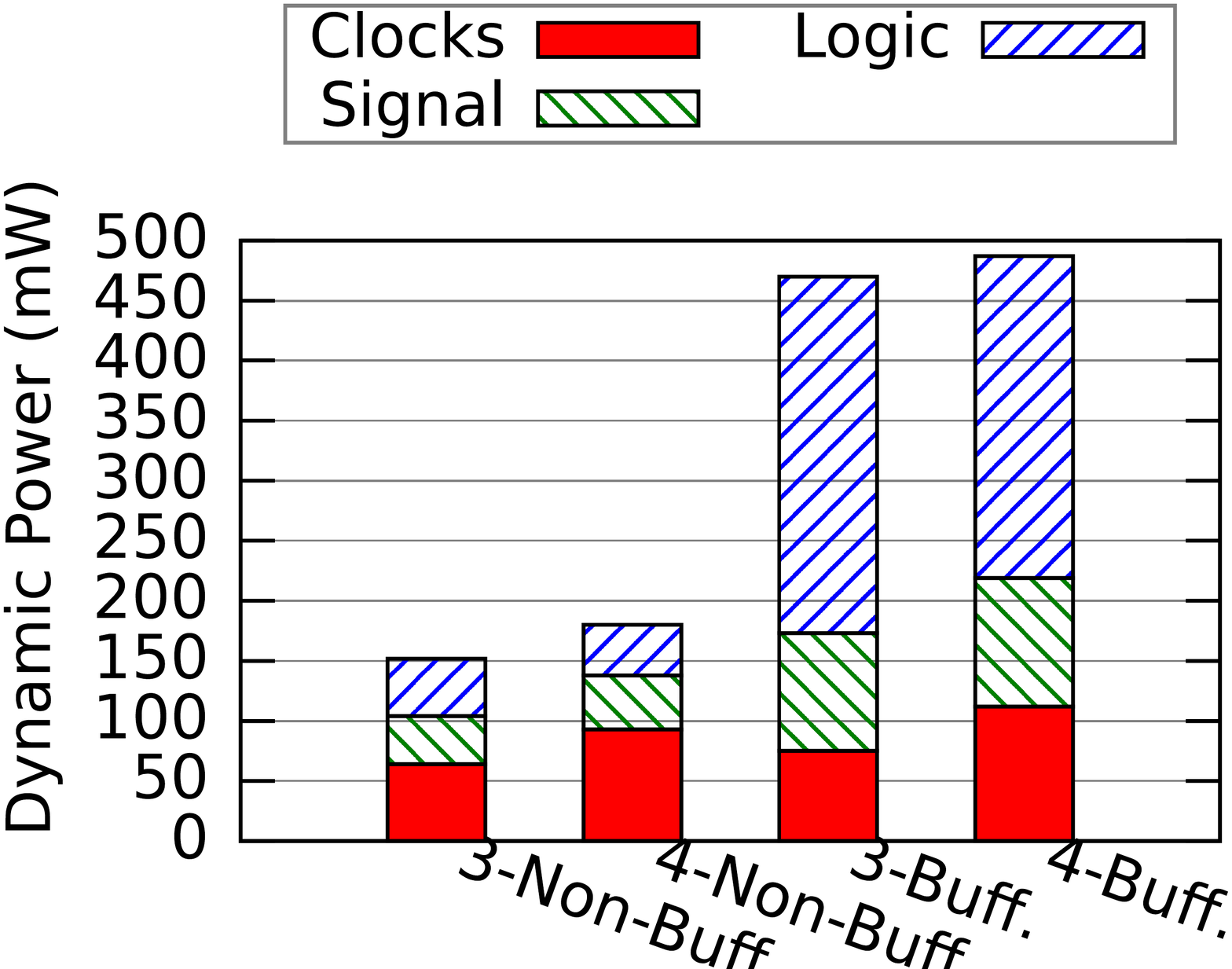}	}
 	\caption{Power Consumption Study of the Routers}
 	\label{fig:powerConsu} 
 	\vspace{-5pt}
\end{figure}

In Figure \ref{fig:resourceUtilization}, we study the resource utilization of the routers. It first evaluates the benefits of optimizing the number of interfaces of the routers. Next, it presents the advantages of removing buffers from the routers. Results are recorded for a data width ranging from 32 bits to 256 bits. The first observation is that reducing the number of port significantly reduces the hardware footprint of the router. In fact, Figure \ref{fig:3_port} and \ref{fig:4_port} show that 3-port routers uses about 40\% less registers and save about 50\% of LUT logic compared to the implementation with 4 interfaces.
The router with buffers even show a more pronounced use of resources with additional LUTs, registers, BRAMs and LUTRAMs (see Figure \ref{fig:3_port_buffer} and \ref{fig:4_port_buffer}). The impact of router resources on power consumption is shown in Figure \ref{fig:powerConsu}.
 First, the  4-port routers that are bufferless can consume up to 2.7$\times$ more power than their 3-port counterparts. Next, buffered routers consume up to 3.11$\times$ more power than bufferless implementations, the highest percentage being recorded from logic. These results demonstrate the benefits in area and power of optimizing the router architecture.

\subsubsection{Maximun Frequency and Latency }
\begin{figure}[]
\vspace{-8pt}
\centering
\includegraphics[width=0.28\textwidth]{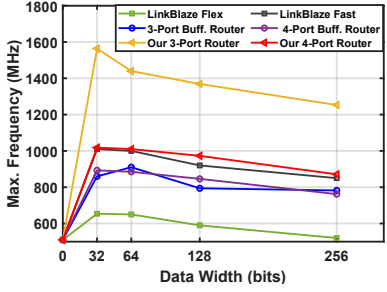}
\caption{Router Scalability Considering the Data Width}
\label{fig:fmax_study} 
\vspace{-6pt}
\end{figure}

\begin{figure}[h]
\vspace{-25pt}
\centering
\includegraphics[width=0.33\textwidth]{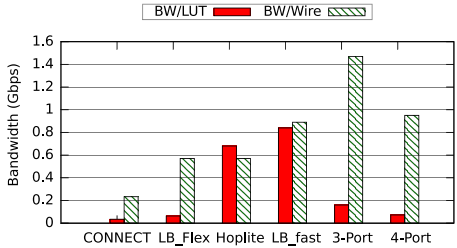}
\vspace{-20pt}
\caption{Bandwidth Comparison to Previous Work}
\label{fig:bandwidth_comp} 
\vspace{-6pt}
\end{figure}

In addition to the area and power benefits, the optimization of the router architecture also results in a higher operating frequency. In Figure \ref{fig:fmax_study}, we compare the maximum frequency of various routers for data width between 32 and 256 bits. We compare our routers to the corresponding buffered implementations, as well as to LinkBlaze Flex and LinkBlaze Fast \cite{maidee2017linkblaze}.
The maximum frequency tends to decrease when the data width increases. This is because larger data widths introduce more logic into the design, which results in additional delays on the data paths. We observe that our routers perform better than the buffered routers and the routers of LinkBlaze Fast/Flex. From the results reported in \cite{maidee2017linkblaze}, CONNECT and Hoplite achieved 313MHz and 638MHz on a Virtex UltraScale+ FPGA. This is far from the 1.5GHz and 1GHz that is achieved respectively by our 3-port and 4-port routers on a similar device. Further, we compare bandwidth results for 32-bit routers to previous work (see Figure \ref{fig:bandwidth_comp}). 
Our 3-port router has 6.3$\times$ better bandwidth per wire than CONNECT, 2.57$\times$ better than Hoplite and LinkBlaze Flex; and 1.65$\times$ better than LinkBlaze Fast. Similar observations can be made for the 4-port router. The bandwidth per LUT nevertheless draws a different picture. Hoplite and LinkBlaze Fast perform better than our routers as they use about 5$\times$ less LUTs than our routers. This is due to the fact that they are less flexible. Hoplite implements a lightweight deflection and is unidirectional, which drastically reduces the size of the routing logic \cite{Kapre2015}. LinkBlaze Fast routers only have 3 ports (2 inputs and 1 output), resulting in lower LUT count \cite{maidee2017linkblaze}.

\begin{figure}[]
\vspace{-8pt}
\captionsetup[subfloat]{farskip=2pt,captionskip=1pt}
	\centering
   
    \subfloat[]{
		\label{fig:latencyStudy}
		\includegraphics[width=.22\textwidth]{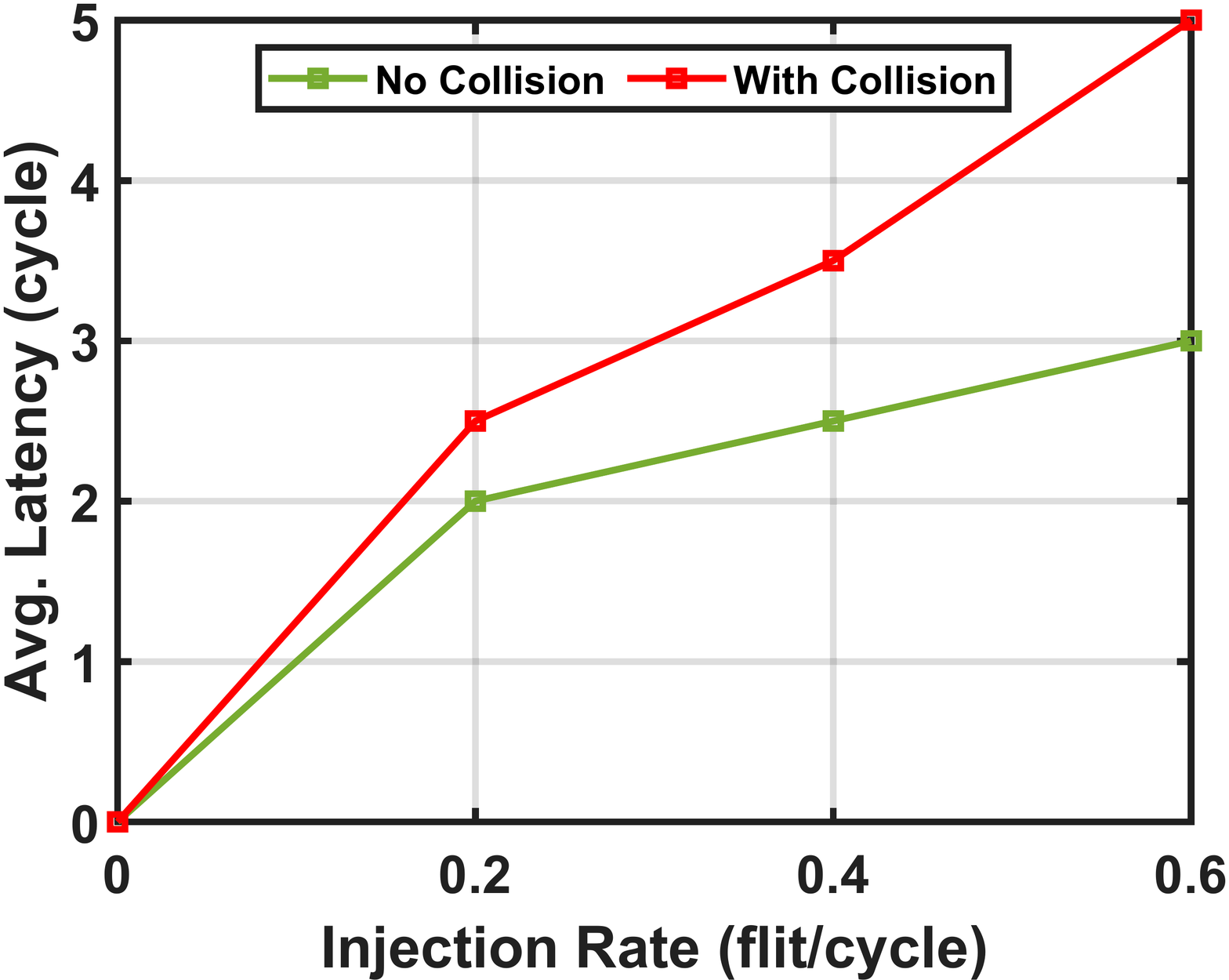}%
	}		
	\hspace{0.0cm}
	\subfloat[]{
		\label{fig:waitingTime}
		\includegraphics[width=.22\textwidth]{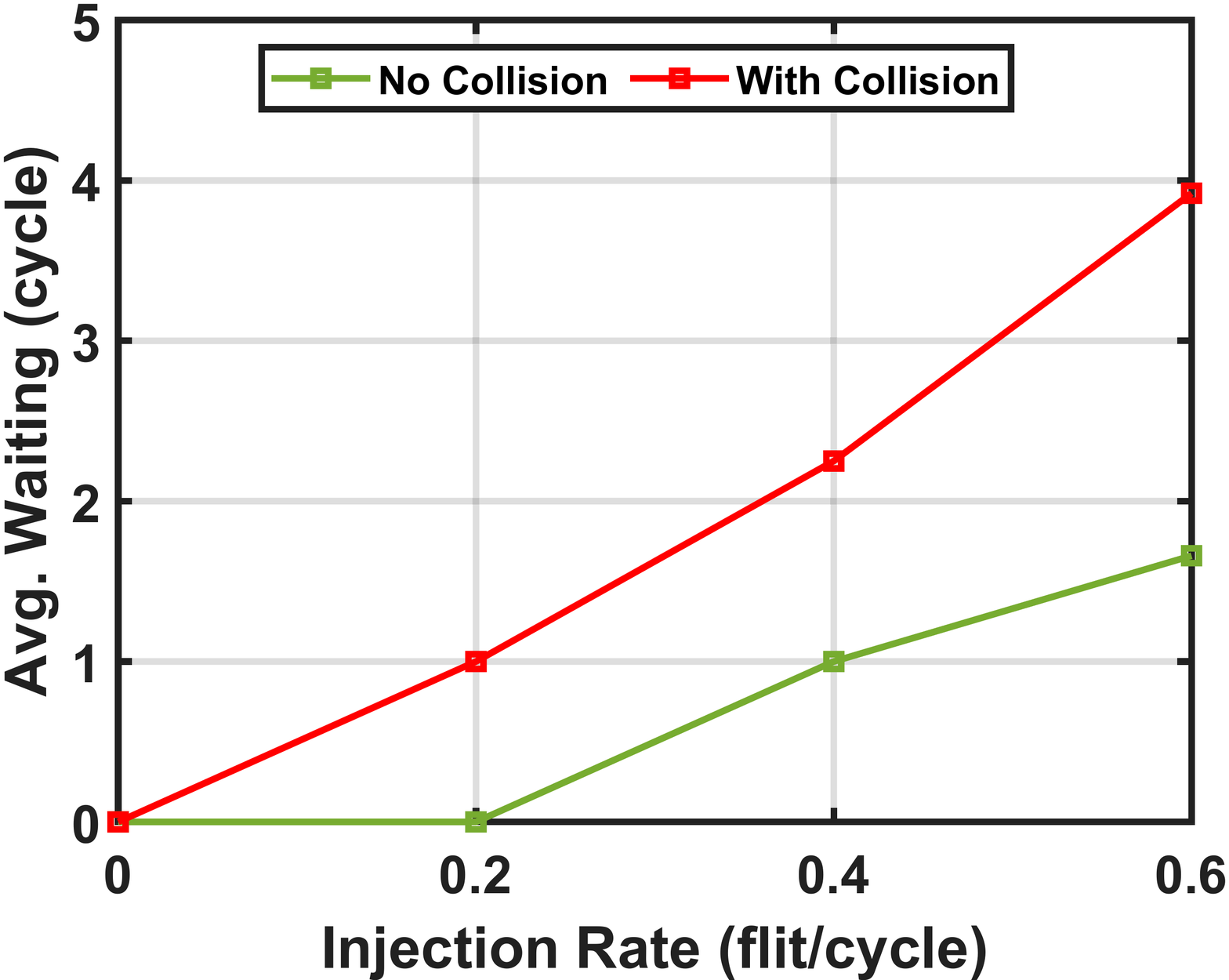}	}

		\caption{(a) Average latency study per injection rate. (b) Average waiting time study per injection rate.}
 	\label{fig:displyaOnSmartphone} 
    \vspace{-6pt}
\end{figure}

We also evaluate our routers against various traffic patterns. Overall, an incoming flit needs two clock cycles to traverse a router. However, when the inputs are pipelined, only the first one will take two cycles. The following packets will be available at the outputs of the router after each cycle (see Figure \ref{fig:collisionManagement}).
Figures \ref{fig:latencyStudy} and \ref{fig:waitingTime} summarize the latency and waiting time observed on our 3-port router in two different configurations. First, we consider when flits arrive from all the interfaces with no collision. In other words, each output port of the router only receives traffic from one input port. With an injection rate of 0.6, the average latency observed is 3 clock cycles and the average waiting is 1.66 clock cycles. Next, we assess the latency and waiting time with collision. In this testing configuration, traffic from two ports target the third port of the router. We observe an increased latency compared to when there is no collision. It is just a consequence of having the packets waiting longer in the VR queues for their turn. In fact, Figure \ref{fig:waitingTime} shows a linear progression of the the waiting curve as the workload increases. The waiting time values when considering collision are about 2$\times$ higher than without collision, which reflects on the average latencies reported in Figure \ref{fig:latencyStudy}.

\subsection{Case Study: FPGA Multi-tenancy}
\label{sec:case_study}

\subsubsection{FPGA division between tenants}
\begin{figure}[h]
\vspace{-8pt}
\centering
\includegraphics[width=0.35\textwidth]{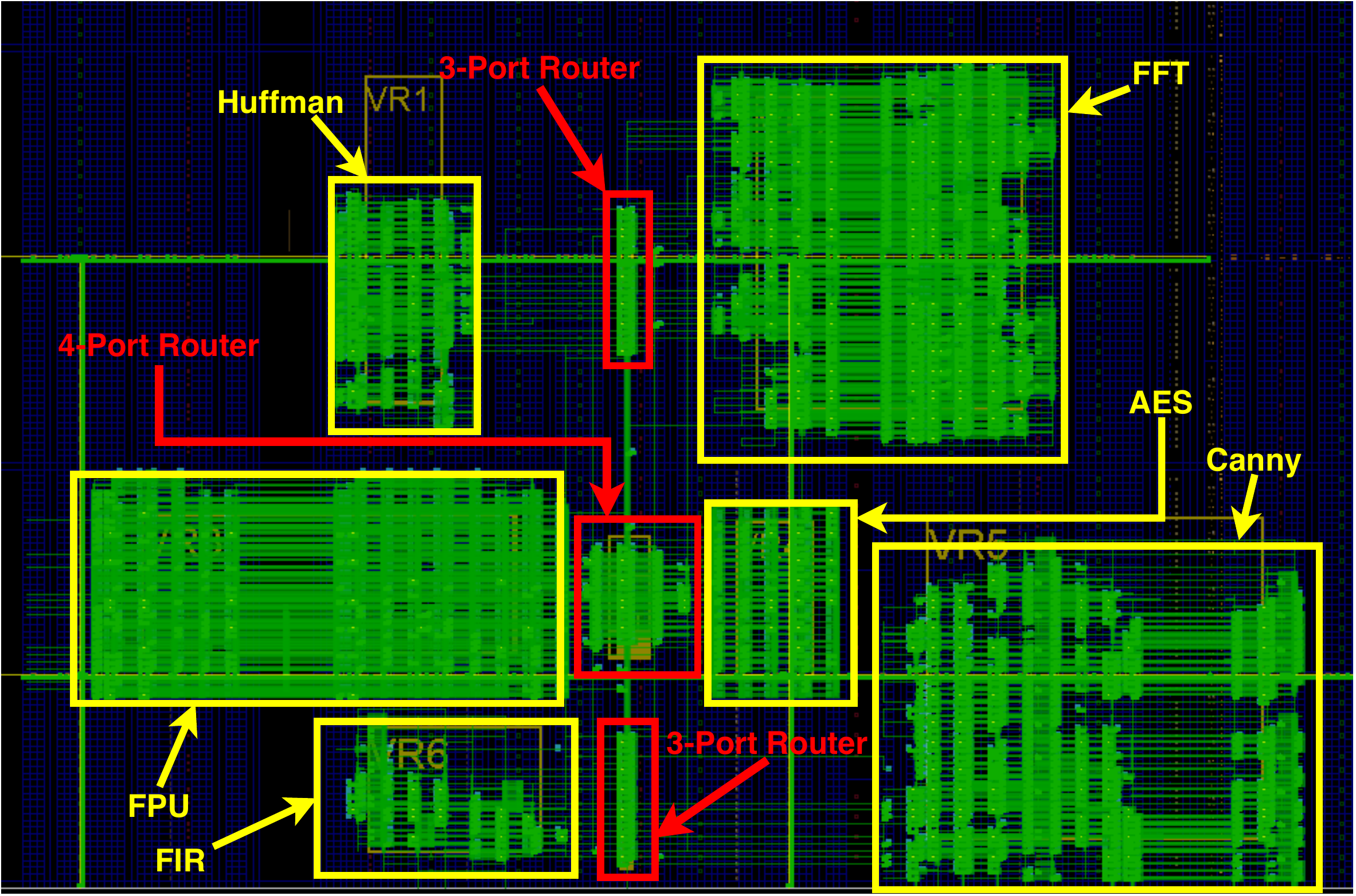}
\caption{Placement of the 6 Jobs from 5 VIs on a single device}
\label{fig:placementJobs} 
\vspace{-6pt}
\end{figure}

To evaluate the FPGA multi-tenancy and elasticity when using our NoC, we consider 5 VIs (labeled VI1,...,VI5) deployed on the OpenStack cloud that access 6 VRs (labeled VR1,...,VR6) on FPGA. The assignment of VRs to VIs is as follows: VR1 is allocated to VI1; VR2 is allocated to VI2; VR3 and VR4 are allocated to VI3; VR5 is allocated to VI4;  VR6 is allocated to VI5. For testing purposes we select 6 hardware accelerators from OpenCores  \cite{opencores}. The applications are: \textit{\textbf{Huffman Decoder}} ---that is typically used in streaming applications; \textit{\textbf{FFT}}---that is heavily used in signal processing; \textit{\textbf{FPU}}---it implements a single precision floating point unit; \textit{\textbf{AES}}---that is an encryption/decryption core over a 128-bit key. \textit{\textbf{Canny Edge}}---implements an edge detection algorithm. \textit{\textbf{FIR}}---is a commonly used filter in signal processing. Table \ref{tab:resource_utilization} summarizes the VR allocation to VIs and the use of resources of test accelerators.
VI3 initially implemented the FPU unit and later requested additional FPGA resource to implement encryption as the two could not fit into the area of VR3. To show the benefits of elasticity with on-chip communication, the FPU streams its output results directly to the AES encryption module through the NoC interconnect. The on-chip communication offers a bandwidth of 25.6 Gbps. Without communication support on the chip, moving data between two VRs will require middleware intervention to copy the data. This could cost around 50$\mu$s (Figure \ref{fig:io_access} reported a minimum of 28$\mu$s for directIO access), which represents a significant performance loss compared to the bandwidth of the NoC. On-chip communication support is therefore of paramount importance to implement efficient hardware elasticity.
\begin{table}[h]
\vspace{-8pt}
\centering
 \tiny
\caption{VR allocation and resource utilization of the applications}
\label{tab:resource_utilization}

\begin{tabular}{|l|l|l|l|l|l|}
\hline
        & \textbf{LUT}  & \textbf{LUTRAM} & \textbf{FF}   & \textbf{DSP} & \textbf{BRAM} \\ \hline \hline
\textbf{Huffman (VR1$\rightarrow$VI1)} & 1288 & 408    & 391  & 0   & 1    \\ \hline
\textbf{FFT (VR2$\rightarrow$VI2)}     & 3533 & 92     & 4818 & 4   & 3    \\ \hline
\textbf{FPU (VR3$\rightarrow$VI3)}     & 4122 & 0      & 582  & 2   & 0    \\ \hline
\textbf{AES (VR4$\rightarrow$VI3)}     & 1272 & 0      & 500  & 0    & 0     \\ \hline
\textbf{Canny Edge (VR5$\rightarrow$VI4)}   & 2558 & 20     & 3825 & 0   & 18   \\ \hline
\textbf{FIR (VR6$\rightarrow$VI5)}     & 270  & 0      & 347  & 4   & 4    \\ \hline

\end{tabular}
\end{table}
For experimental purposes, we implement the \textit{single-column} division of the FPGA (see section \ref{subsec:topology}). Since we have 6 VRs, we will only need 3 routers (two 3-port routers and one 4-port router). The routers support 32-bits datapaths. Figure \ref{fig:placementJobs} shows a screenshot of area occupied by the NoC and each of the applications. For the sake of brevity, we do not show the rest of the shell that controls IO interfaces for off-chip communication. Because of the high capacity of integration of high-end FPGA architectures such as the UltraScale+ family, the size of each VR could easily be close to that of an entire legacy FPGA. For instance, the pblock defining VR5 occupies 1121 CLBs, or 8968 LUTs (0.22\% of the LUTs in VU9P) which represents about 20\% of some FPGAs from the 7-series \cite{7series}. This means that while a device from the 7-series may only be able to host about 5 instances of size equal to VR5, a VU9P device could deploy about 455 instances of those. 
This observation highlights the need for spatial sharing support to increase device utilization. The NoC and applications illustrated in Figure \ref{fig:placementJobs} only used 1.71\% of the CLB area of the FPGA, leaving enough room for additional workloads from cloud users. 
The 3-port and 4-port routers respectively cover  305 LUTs (0.03 \% of the FPGA) and 491 LUTs (0.04\% of the chip).

\subsubsection{IO Trip and Throughput Study}
\begin{wrapfigure}[9]{r}{0.3\textwidth}
\vspace{-23pt}
\centering
\includegraphics[width=0.3\textwidth]{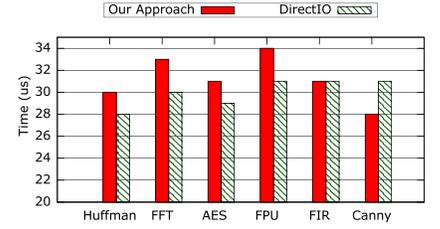}
\vspace{-30pt}
\caption{IO Trip Comparison}
\label{fig:io_access} 
\vspace{-8pt}
\end{wrapfigure}

first, we measure the overhead introduced by the cloud management software on the FPGA access time. We want to compare the IO performance in multi-tenant and single-tenant deployments to show that the spatial sharing of FPGAs does not significantly affect the QoS. We then consider two modes: \textbf{(1) Multi-tenant (Our approach):} all the 6 applications are deployed as illustrated in Figure \ref{fig:placementJobs}. The VIs continuously write, then read from the accelerators and we record the IO trip times. \textbf{(2) Single-tenant (DirectIO):} The entire FPGA is successively allocated to each VI that runs write, then read operations and we record IO trip times. Figure \ref{fig:io_access} summarizes the average IO trip recorded time. It is observed that there is no significant difference in IO cost between the two schemes as they both simply consist in accessing FPGA registers from the host/guest operating systems. 
An IO access time penalty is however recorded when requests arrive simultaneously from different tenants at the entry point of the shared device. Such requests are queued in the cloud management software and the IO access delays observed are only in the order of a few microseconds. As example, an IO round trip to the AES core takes in average $31\mu s$ in the multi-tenant deployment while using about $29\mu s$ in the single-tenant FPGA allocation. On the other hand, accessing the FIR IP took in both cases an average of $31\mu s$. There are also cases where the IO requests performed better in the multi-tenant configuration. This means that we have achieved a 6$\times$ higher FPGA utilization rate as a single device is transparently running 6 different workloads. It is worth to note that these results were recorded in a configuration in which the FPGA was connected to the same physical server running the VIs (the FPGA node was purposely merged with the all-in-one OpenStack node for fair comparison with the directIO scheme). As we will discuss later, remotely accessing the FPGA incurs network transmission overhead. 

\begin{figure}[h]
\vspace{-8pt}
\captionsetup[subfloat]{farskip=2pt,captionskip=1pt}
	\centering
   
    \subfloat[]{
		\label{fig:sameServer}
		\includegraphics[width=.225\textwidth]{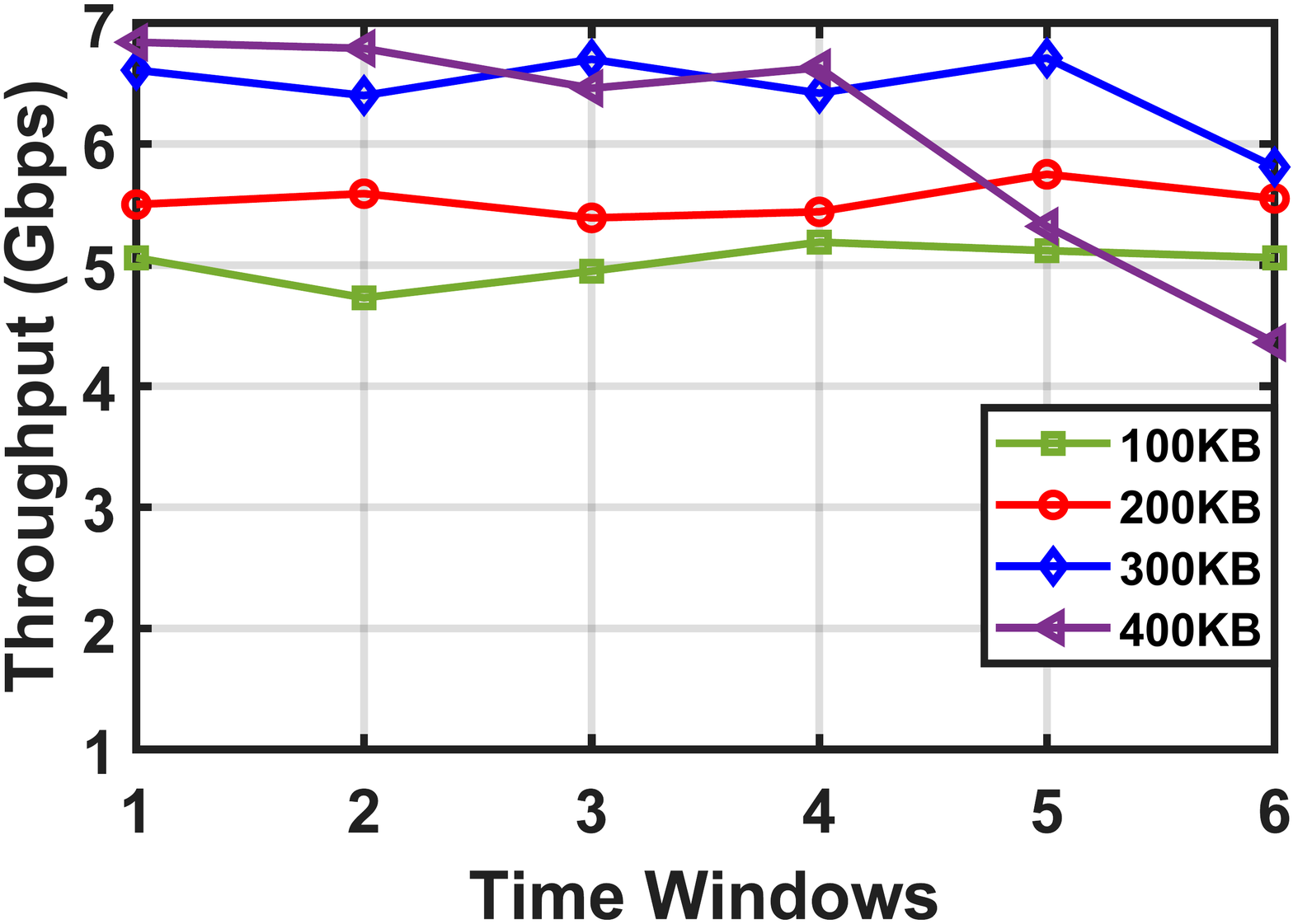}%
	}		
	\hspace{0.1cm}
	\subfloat[]{
		\label{fig:distantServer}
		\includegraphics[width=.22\textwidth]{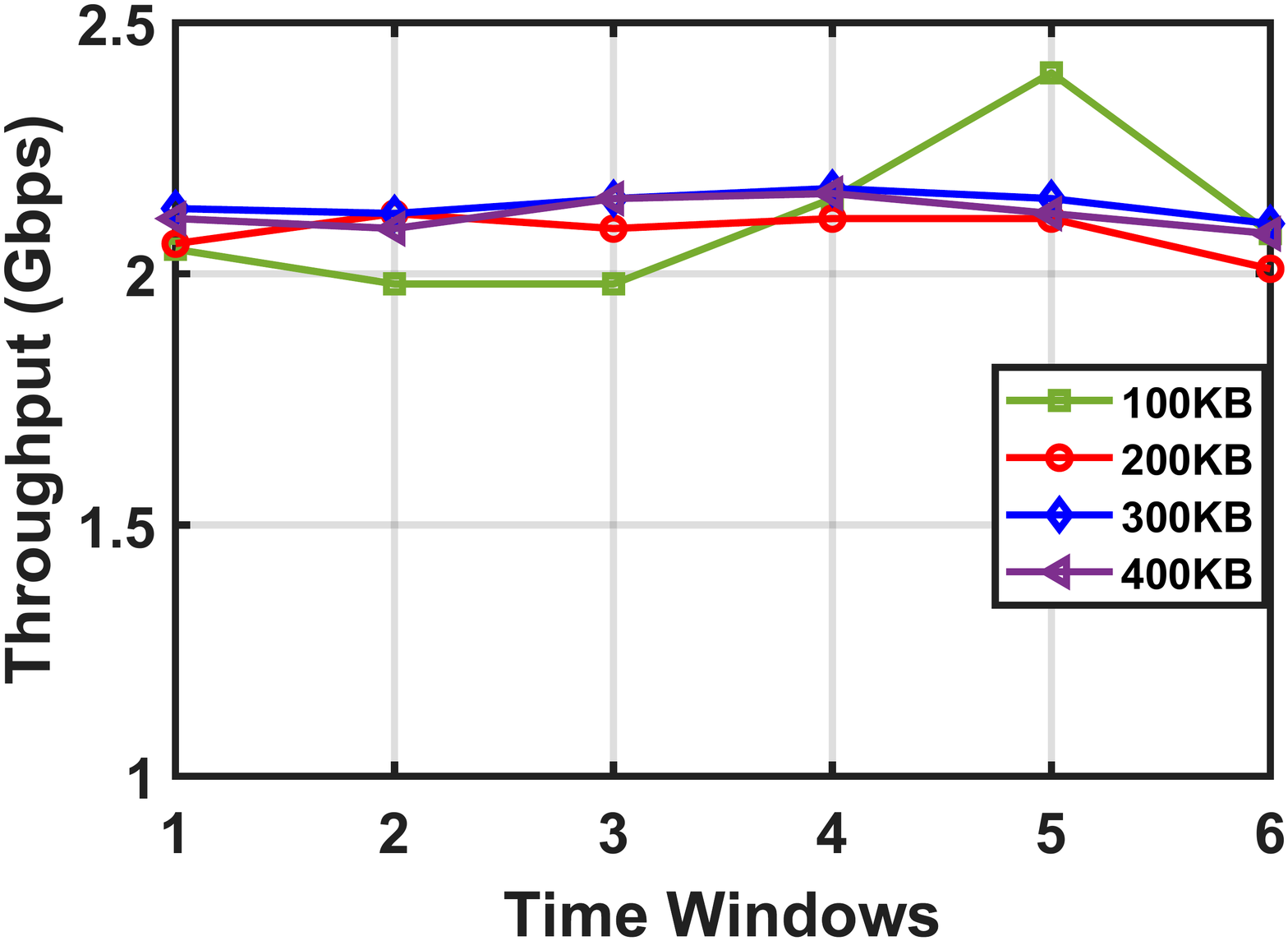}	}

		\caption{(a) Throughput Study when the Virtual Instances are deployed on the physical Server hosting the FPGA devices. (b) Throughput Study when the Virtual Instances access FPGAs remotely.}
 	\label{fig:displyaOnSmartphone} 
 	\vspace{-6pt}
\end{figure}

We also study the throughput achieved on the multi-tenant cloud FPGAs. We continuously stream packets of size ranging from 100KB to 400KB between the VIs and hardware accelerators on FPGA, and record the average of throughput observed. Throughput data is collected over an hour of operation after 6 random time windows with all the VIs deployed on the server hosting the FPGA (Figure \ref{fig:sameServer}) and with VIs remotely accessing the FPGA node over the Ethernet (Figure \ref{fig:distantServer}).

When VIs access FPGA accelerators hosted on the same physical server, we observe a throughput reaching 7Gbps for 400KB payloads, which is about $2\times$ higher than the software to hardware and hardware to software throughput reported in \cite{eskandari2019modular}. Up to $3\times$ performance lost is however observed in distant FPGA access as the throughput is limited by the bandwidth of the Ethernet router (see section \ref{subsec:platform}). 

\subsubsection{Comparison with previous work}
 Table \ref{tab:architecture_comparison} puts into perspective our proposed architecture compared to other reported FPGA enabled cloud schemes.  As the table shows, DirectIO provides better performance compared to our approach but does not offer actual virtualization benefits such as resource re-allocation at runtime. Our approach appears as the best trade-off as it enables runtime re-allocation, hardware elasticity, local communication between VRs hosted on the same device. The work presented in \cite{mbongue2018fpga} has a lower IO trip time, but is technology-specific as it only works for KVM clouds.

\begin{table}[h]
\vspace{-10pt}
\centering
 \tiny
\caption{Cloud FPGA Architecture Comparison}
\label{tab:architecture_comparison}
\begin{tabular}{|c|c|c|c|c|}
\hline 
\textbf{Works} & \textbf{\begin{tabular}[c]{@{}c@{}}Runtime \\ Re-allocation \\ Support\end{tabular}} & \textbf{\begin{tabular}[c]{@{}c@{}}Hardware \\ Elasticity\\ Support\end{tabular}} & \textbf{\begin{tabular}[c]{@{}c@{}}On-Chip \\  Com. \\Support\end{tabular}} & \textbf{\begin{tabular}[c]{@{}c@{}}IO Trip\\  Cost\\  (in $\mu s$)\end{tabular}} \\ \hline \hline
DirectIO & No & Yes & Yes & 28 \\ \hline
\textbf{Our Work} & \textbf{Yes} & \textbf{Yes} & \textbf{Yes} & \textbf{30} \\ \hline
\cite{chen2014enabling} & Yes & No & No & 15 \\ \hline
\cite{byma2014fpgas} & Yes & No & No & 600 \\ \hline
\cite{mbongue2018fpga} & Yes & Yes & Yes & 26 \\ \hline

\cite{vaishnav2018resource} & Yes & Yes & No & -- \\ \hline

\cite{asiatici2017virtualized} & Yes & No & No & 8000 \\ \hline
\cite{fahmy2015virtualized} & Yes & No & No & 16000 \\ \hline
\end{tabular}
\vspace{-8pt}
\end{table}



\section{Conclusion}
\label{sec:conclusion}
This work proposed an approach to enable spatial sharing of FPGA resource between multiple tenants in the cloud. We leverage a NoC architecture to implement elasticity. In the context of this work, we considered the elasticity as the ability to assign additional FPGA components to users at run-time. The proposed NoC makes it possible assign multiple FPGA regions to users and implement fast data movement to support on-chip communication between running workloads. Experiments demonstrated the low resource utilization and high frequency of operation of our architecture, as well as an increased FPGA utilization.

\section*{Acknowledgement}
This work was partially supported by the ONR under the Grant CCN 0402-17643-21-0000, and the Air Force Research Lab AFRL/RIGA Cyber Assurance Branch, Rome NY.


\bibliographystyle{./IEEEtran}
\bibliography{fpl-2019-ds-overlay}


\end{document}